# MEASURING BANDWIDTH FOR SUPER COMPUTER WORKLOADS

A.Neela madheswari[1], Dr.R.S.D.Wahida banu[2],
[1]Research Scholar, Anna University, Coimbatore.
[2]Research Supervisor, Anna University, Coimbatore.

*Abstract*

Parallel computing plays a major role in almost all the fields from research to major concern problem solving purposes. Many researches are till now focusing towards the area of parallel processing. Nowadays it extends its usage towards the end user application such as GPU as well as multi-core processor development. The bandwidth measurement is essential for resource management and for studying the various performance factors of the existing super computer systems which will be helpful for better system utilization since super computers are very few and their resources should be properly utilized. In this paper the real workload trace of one of the super computers LANL is taken and shown how the bandwidth is estimated with the given parameters.

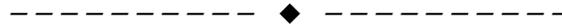

## 1. Introduction

Measuring the available bandwidth is of great importance for predicting the end-to-end performance of applications, for dynamic path selection and traffic engineering, and for selecting between a numbers of differentiated classes of service. The available bandwidth is an important metric for several applications, such as grid, video and voice streaming, overlay routing, p2p file transfers, server selection, and inter-domain path monitoring [8].

The performance of a computer system depends not only on its design and implementation, but also on the workload to which it is subjected [3]. During the design process of a parallel computer and its management software, the evaluation of the system is an important task. Here, the availability of appropriate workloads is necessary for quantitative evaluations. Substantial information is required about the workload which is executed on these systems. Ideally, the exact workload is available and can be used during the design process. However, usually the exact workload for a certain system is not known during the design process [6].

## 2. Background

The price of supercomputers is dropping quickly, in part because they are often built with the same off-the-shelf parts found in PCs. Just about any organization with a few million dollars can now buy or assemble a top-flight machine. But over the last 10 years, the vital innards of supercomputers have become more main stream and a wide variety of organizations have bought them. A supercomputer called Jaguar at the Oak Ridge National Laboratory in Tennessee officially became the world's fastest machine. It links thousands of mainstream chips from Advanced Micro Devices [5].

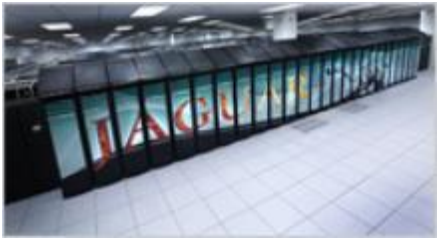

Fig.1 Jaguar supercomputer at the Oak Ridge National Laboratory, the world's fastest, links thousands of mainstream chips.

There are many researches done based on the workload of supercomputing environment. They are focusing towards how to model several workload parameters such as inter-arrival time, job response time, required amount of memory, etc. But for the actual analysis the remaining informations may be left out.

In [4], a model is presented to recover an estimated job execution time when this information is not available. Here the real workload traces are used. Alternatively workload models are also applied.

In [7], they provide a summary and classification of the most common techniques used for characterizing several types of information and commutation systems. They proposed a methodology for workload characterization framework. They provide general guidelines for deriving a good workload model suitable as an input to performance studies.

In [2], they have characterized the inter-arrival time and service time distributions for jobs at a large MPP supercomputing center. They have presented the parameters from the characterization so that they can be easily used for both theoretical studies and the simulations of various scheduling algorithms.

## 3. Workload availability

A few years ago practically no real data about production workloads on parallel machines are available, so evaluations had to rely on guess work. This situation is changed dramatically and now practically all evaluations of supercomputer workloads rely on real data at least to some degree. There are two forms of workloads are available.

### 3.1. Workload logs

Most parallel super computers maintain accounting logs for administrative use. These logs contain valuable information about all the activity executed. The format of the log is mostly ascii file with one line per job. Analyzing such logs can lead to important insights into the workload. Such work has been done for some systems such as CTC SP2 [14] and llnlthunder [15], etc. The large collections of parallel log files are available in [10].

### 3.2. Workload models

Workload models are based on some statistical analysis of workload logs with the goal of elucidating their underlying principles. This enables the creation of new workloads that are statistically similar to the observations but can also be changed at will.

The workload models fall into two categories namely, rigid jobs and fixed jobs. Rigid job model create a sequence of jobs with given arrival time, number of processors and runtime. The task of the scheduler is then to pack these rectangular jobs onto the machine. Flexible job models provide data about the total computation and the speed function, instead of the required number of processors and the runtime.

A model can be invalid, that is not the representative of real workloads in many different ways. The most obvious is using the wrong distributions. To be fair finding the right distribution is not always easy and there are various methodological options that do not necessarily produces the same results. If a workload model is used, one must ensure that it is a good one [5].

Since the real workload trace is available from [10], in this paper, the traces are make use off to find the bandwidth for the existing supercomputer workload evaluation.

## 4. Bandwidth estimation

The workload from the supercomputer of Los Alamos National Lab. The system consists of 1024-node connection Machine CM-5 from Thinking Machines. The total trace that is taken during Oct 1994 to Sep 1996 are taken as a whole of totally 201387 jobs. For more information regarding this centre, refer [11].

The required data needed for measuring bandwidth are:
1. Number of bytes needed for execution – n.
2. Start time of the execution of the job – stdate.
3. End time of the execution of the job – enddate.

Then the bandwidth is measured for each job as,

$$B = 1000 \times n / diff$$

where diff = enddate – stdate.

For the workload of LANL, the various fields available are:
1. JobId.
2. Submit date and time.
3. Start date and time.
4. End date and time.
5. Requested processors.
6. Used processors.
7. Requested CPU time in secs.
8. Used CPU time in secs.
9. Requested memory in Kbytes.
10. Used memory in Kbytes.
11. Queue.
12. Dedicated.
13. User.
14. Project.
15. Executable.
16. Exit code.

But for estimating the bandwidth, the data needed are taken as,
For n, the values from the field 9 is taken.
For stdate, field 3 value is considered and for the enddate, field 4 value is considered.

The sample coding for finding the bandwidth is given by:

```
public float getRateFor(int sidx)
{
float rate = 0;
int scnt = samples.size();
if (scnt > sidx && sidx >= 0)
{
DataSample s =
(DataSample)samples.elementAt(sidx);
Date start = s.start;
Date end = s.end;
if (start == null && sidx >= 1)
{
DataSample prev =
(DataSample)samples.elementAt(sidx - 1);
start = prev.end;
}
```

```
if (start != null && end != null)
{
long msec = end.getTime() - start.getTime();
rate = 1000 * (float)s.byteCount /
(float)msec;
}
}
return rate;
}
```

Out of 201387 jobs, the availability of data is for 135375 jobs only since some of the start and end dates are unknown for fewer jobs. So those jobs are omitted and then the bandwidth measurement is considered for the remaining 135375 jobs. Finally the bandwidth is measured for all the jobs in terms of Mbytes/sec.

Similarly where we are having the information regarding the required amount of memory to execute the job with start and end time are available, we can able to estimate the bandwidth using the specified method. We can able to extend to find the bandwidth for the jobs that are running under CTC [12], SDSC Blue Horizon, Sharcnet [13].

The sample worksheet is given in the table below. Here the start date and end date denotes the start and end date of a particular job and Mbytes is the estimated bandwidth for that particular job and Bytes are the required bytes which are available in the workload trace itself.

Similarly we can able to estimate for the remaining workloads given from the parallel workload archives given at [10]. This work can be used for the paper [9] also as a future extension for the existing parallel workloads since there they have given only for the data of workload models.

**Table1.** Sample worksheet for estimated bandwidth output

| Start date | End date | Mbytes | Bytes |
|---|---|---|---|
| May 10 94 | May 10 94 | 204.7445 | 16777216 |
| May 10 94 | May 10 94 | 9.746579 | 755712 |
| May 10 94 | May 10 94 | -2.78E-04 | 32768 |
| May 10 94 | May 10 94 | 124.6248 | 16777216 |
| May 10 94 | May 10 94 | 68478.43 | 122880 |
| May 10 94 | May 10 94 | 84.90879 | 16777216 |
| May 10 94 | May 10 94 | 463.3484 | 204800 |
| May 10 94 | May 10 94 | 682.6667 | 512000 |
| May 10 94 | May 10 94 | 134.7812 | 32768 |
| May 10 94 | May 10 94 | 1424.696 | 409600 |
| May 10 94 | May 10 94 | 1024 | 32768 |
| May 10 94 | May 10 94 | 1424.696 | 307200 |
| May 10 94 | May 10 94 | 1365.333 | 32768 |
| May 10 94 | May 10 94 | 2520.616 | 32768 |
| May 10 94 | May 10 94 | 606.8148 | 32768 |
| May 10 94 | May 10 94 | 268.5902 | 32768 |

## 5. Conclusion

Evaluation methods for parallel computers often require the availability of relevant workload information. Workload traces recorded on real super computing environment is used here. The presented method is used to find the bandwidth if it is not available in the workload traces which can then be used to analyze various metrics of how to optimize the performance of the super computing environments further.

**References**

[1] Java Distributed Computing, Oreilly.

[2] Joefon Jann, Pratap Pattnaik, Hubertus Franke, Fang Wang, Joseph Skovira and Joseph Riodan, "Modeling of workloads in MPPs", In the Job Scheduling Strategies for Parallel Processing, D.G.Feitelson and L.Rudolph (Eds.) Springer-Verlag, 1997, LNCS vol 1291, pp. 95-116.